\documentclass[conference]{IEEEtran}
\usepackage{cite}
\usepackage{amsmath,amssymb,amsfonts}
\usepackage{graphicx}
\usepackage{textcomp}
\usepackage[dvipsnames]{xcolor}
\usepackage{xspace}
\usepackage{tabularx, booktabs}
\usepackage{algorithm}
\usepackage{algpseudocode}
\usepackage{pifont}
\usepackage{diagbox}
\usepackage{multirow}
\usepackage{makecell}
\usepackage{enumitem}
\def\BibTeX{{\rm B\kern-.05em{\sc i\kern-.025em b}\kern-.08em
    T\kern-.1667em\lower.7ex\hbox{E}\kern-.125emX}}

\newcommand{\tbhl}[1]{{#1$^\dagger$}}

\newcommand{\ouralg}{PermuteV\xspace}
\newcommand{\xmark}{\ding{55}\xspace}
\newcolumntype{Y}{>{\centering\arraybackslash}X}
\newcolumntype{Z}{>{\quad}X}
    
\begin{document}

\title{\ouralg: A Performant Side-channel-Resistant RISC-V Core Securing Edge AI Inference}

\author{
\IEEEauthorblockN{Nuntipat Narkthong\IEEEauthorrefmark{1}, Chattriya Jariyavajee\IEEEauthorrefmark{2}, Xiaolin Xu\IEEEauthorrefmark{1}}
\IEEEauthorblockA{\IEEEauthorrefmark{1}Northeastern University, Boston, USA \IEEEauthorrefmark{2}King Mongkut's University of Technology Thonburi, Bangkok, Thailand
    \\\{narkthong.n, x.xu\}@northeastern.edu chattriya.jar@mail.kmutt.ac.th}
}

\author{\IEEEauthorblockN{Nuntipat Narkthong}
\IEEEauthorblockA{
\textit{Northeastern University}\\
Boston, USA\\
narkthong.n@northeastern.edu}
\and
\IEEEauthorblockN{Xiaolin Xu}
\IEEEauthorblockA{
\textit{Northeastern University}\\
Boston, USA\\
x.xu@northeastern.edu}
}
\maketitle

\begin{abstract}
Edge AI inference is becoming prevalent thanks to the emergence of small yet high-performance microprocessors. This shift from cloud to edge processing brings several benefits in terms of energy savings, improved latency, and increased privacy. On the downside, bringing computation to the edge makes them more vulnerable to physical side-channel attacks (SCA), which aim to extract the confidentiality of neural network models, e.g., architecture and weight. To address this growing threat, we propose \ouralg, a performant side-channel resistant RISC-V core designed to secure neural network inference. \ouralg employs a hardware-accelerated defense mechanism that randomly permutes the execution order of loop iterations, thereby obfuscating the electromagnetic (EM) signature associated with sensitive operations. We implement \ouralg on FPGA and perform evaluations in terms of side-channel security, hardware area, and runtime overhead. The experimental results demonstrate that \ouralg can effectively defend against EM SCA with minimal area and runtime overhead.
\end{abstract}


\section{Introduction}

Edge devices like smart home and IoT systems have become much more powerful recently, allowing them to perform complex tasks like neural network inference locally. This shift from cloud-based processing to edge computing offers numerous benefits, including reduced latency, lower power consumption, and enhanced user privacy. However, performing computation directly on these devices makes them more susceptible to side-channel attacks (SCAs), which could allow attackers to steal the intellectual property stored on the devices, such as the weight and architecture of the neural network model. 

Side-channel attacks exploit the physical characteristics of hardware devices such as power consumption and electromagnetic emanations, to extract sensitive information. Since the pioneering work of Kocher et al.~\cite{kocher1999differential}, SCAs have been extensively studied, initially focusing on cryptographic algorithm implementations \cite{ches-2004-625, 1286711, 7357115}. Recently, researchers have demonstrated the feasibility of applying SCAs to neural networks, extracting model architectures and weights through side-channel analysis~\cite{Batina2019,Takatoi2020,ShuffleV}.

Exposure of neural network (NN) models to SCAs poses significant risks. High-performance NN models, developed through extensive training and substantial resource investment, represent valuable intellectual property of the vendor. More importantly, the model's internal architecture and parameters, if leaked, can aid attackers in performing adversarial attacks \cite{Dalvi2004} or inferring sensitive training data \cite{carlini2021extracting}. Therefore, protecting these models from reverse engineering is a vital task.

Several defense mechanisms targeting NN hardware accelerators have been proposed based on shuffling and masking techniques\cite{Dubey2020_ICCAD, Dubey2020, luo2022nnrearch, Liu2019}. While these methods achieve low runtime overhead, they often introduce substantial area overhead (up to 5.9x) and are not applicable to software-based NN implementation running on general-purpose microprocessors which are common way to deploy NN on edge devices \cite{10.1145/3661820,10.1145/3620665.3640374,10.5555/3495724.3496706}. Conversely, microarchitecture-level protections for microprocessors such as \cite{Bayrak2012,Antognazza2021} are only designed for and evaluated on cryptographic encryption workload. The only existing protection targeting NN workload on microprocessors is \cite{ShuffleV}. While effective on NN workloads and requiring no software modification, it incur significant runtime overhead (13.7\% to 41.8\%). This overhead hinders their adoption in resource-constrained edge devices with strict latency requirements. Therefore, there is a clear need for effective countermeasures that can safeguard NN models on edge devices powered by general-purpose microprocessors with minimal overhead.

To address this pressing concern, this paper introduces \ouralg, a side-channel resistant RISC-V core design to secure neural network inference on edge devices against physical SCAs. We developed \ouralg on an open-source Ibex RISC-V core \cite{ibex} to facilitate its adoption in real-world system on a chip (SoC) design. \ouralg incorporates a Loop Index Generator (LIG) module and custom instruction set architecture (ISA) extensions to permute the execution order of loop iterations with negligible performance overhead. To facilitate software development, we extended the LLVM compiler framework \cite{lattner2004llvm} to automatically generate code utilizing these custom ISA extensions. Our evaluation demonstrates that \ouralg's permutation strategy effectively obfuscates electromagnetic emanations associated with sensitive computations, significantly hindering attackers' ability to extract valuable information.

We make the following contributions in this work.

\begin{enumerate}
    \item We present \ouralg, the first microarchitecture and software co-designed solution for secure neural network inference on general-purpose microprocessors. \ouralg integrates a Loop Index Generator (LIG) module to permute the loop iteration order, thwarting EM SCAs with negligible runtime overhead without requiring code modifications. 
    
    \item We propose RISC-V ISA extensions to leverage the LIG module. \ouralg custom instructions implicitly add the permuted loop iteration index to the source register value, enabling efficient hardware implementation and allowing the compiler to generate permuted loop code using the same number of instructions as in a sequential loop. 

    \item We implement \ouralg on an FPGA and evaluate its side-channel security and performance on a multiply-accumulate (MAC) operation, which is the building block of the NN inference operation. \ouralg is fully compatible interface-wise with the popular open-source Ibex core \cite{ibex}, enabling its real-world use as a drop-in replacement core in many existing SoC design projects like PULP \cite{PULP} and OpenTitan \cite{opentitan}. 
\end{enumerate}

\section{Background and Related Works}

\subsection{EM Side-Channel Attack against Neural Networks}

Side-channel attacks (SCAs) represent a significant threat to the security of various systems, including but not limited to cryptosystems. Unlike traditional attacks that exploit software vulnerabilities, SCAs target the physical characteristics of devices such as power consumption \cite{kocher1999differential}, electromagnetic (EM) emanations \cite{Batina2019,Takatoi2020}, and timing information \cite{Gerlach2023} to extract sensitive information.

Correlation Electromagnetic Analysis (CEMA) is a particularly effective SCA technique that leverages the correlation between EM emanations and a device's internal state. By collecting and analyzing EM traces, attackers can infer confidential information, such as cryptographic keys. A typical CEMA attack begins with the construction of a behavioral model of the target device, incorporating potential secret values. This model is then used to predict the device's EM emissions for different candidate secret values. Comparing these predictions with actual measurements enables attackers to identify the most likely secret.

Although extensively studied in the context of cryptographic systems \cite{10097781}, recent research has shown that CEMA can also be applied to attack hardware and software implementations of neural network models \cite{Batina2019,Takatoi2020,Cheng2024,ShuffleV}. In such attacks, the attackers can potentially extract critical information, such as the model's architecture and weights, by analyzing the EM emanations from a device performing neural network inference.

\subsection{Defense against EM Side-Channel Attacks}
To mitigate these threats, researchers have introduced various defense mechanisms to protect both NN architecture and its weight, primarily focusing on hardware NN accelerators. For instance, Liu et al.~\cite{Liu2019} introduced a technique to safeguard DNN model architecture by shuffling and adding dummy memory accesses. Luo et al.~\cite{luo2022nnrearch} further extended this approach to protect DNN architectures on the open-source FPGA-based DNN accelerator, VTA \cite{vta}, using advanced scheduling obfuscation methods. Additionally, Dubey et al.~\cite{Dubey2020,Dubey2020_ICCAD} proposed a fully masked neural network inference engine on an FPGA that can protect both model architecture and model weight with 3.5\% overhead in terms of latency and a 5.9x increase in terms of area. While these efforts effectively protect NN hardware implementation, they are not applicable to software-based NNs running on general-purpose microprocessors.

On microprocessor, several defense techniques including shuffling, code morphing, and masking have been proposed to mitigate side-channel attacks on the microarchitectural level. For example, Bayrak et al.~\cite{Bayrak2012} introduced a hardware-based approach to randomize the execution order of independent instruction blocks based on manual or compiler-assisted code annotation. 
Antognazza et al.~\cite{Antognazza2021} proposed Metis, a hardware module that performs code morphing at the microarchitecture level, resulting in lower runtime overhead compared to purely software-based morphing approaches. Although Metis successfully protects diverse cryptographic algorithms, it still incurs high runtime overhead due to the inherent nature of code morphing, which relies on converting one instruction into multiple functionally equivalent instructions.
Alternative, Gross et al.~\cite{Hannes2017} applied domain-oriented masking technique \cite{10.1145/2996366.2996426} to secure the open-source V-scale RISC-V core against passive physical attacks. While demonstrating effective protection for cryptographic applications like ASCON~\cite{ASCON}, this approach leads to considerable hardware overhead, requiring at least 1.59x more LUTs and 1.84x more registers on a Xilinx Spartan-6 FPGA.
One common limitation of these work is the lack of evaluation on neural-network workload. While these defenses for general-purpose microprocessors may be applicable to protect NN inference, their suitability and performance characteristics under the neural network workloads remain unexplored. 

The only microarchitectural-level defense on general-purpose microprocessors that has been designed for and evaluated on neural network workloads is ShuffleV \cite{ShuffleV}, which proposes integrating a hardware module to the RISC-V core to randomly shuffle the execution order of program instructions and insert dummy instructions automatically without any software modification or recompilation. Although it successfully protects both AES encryption and neural network workloads, it still incurs a high runtime overhead (13.7\% to 41.8\%) and an area overhead of 22.7\% compared to the unprotected core. Our work aims to address this critical gap by proposing a RISC-V ISA extension along with a hardware module specifically designed for and evaluated with an NN workload, which can achieve negligible runtime overhead to suit resource-constrained edge devices. The main difference between our approach and ShuffleV is that we use a modified compiler to perform code transformations and emit custom instructions, allowing us to utilize our hardware module during compile time. Therefore, \ouralg eliminates the high runtime overhead associated with performing transformations and analyses on the hardware.

\begin{figure*}[]
    \centering
    \includegraphics[width=0.95\linewidth]{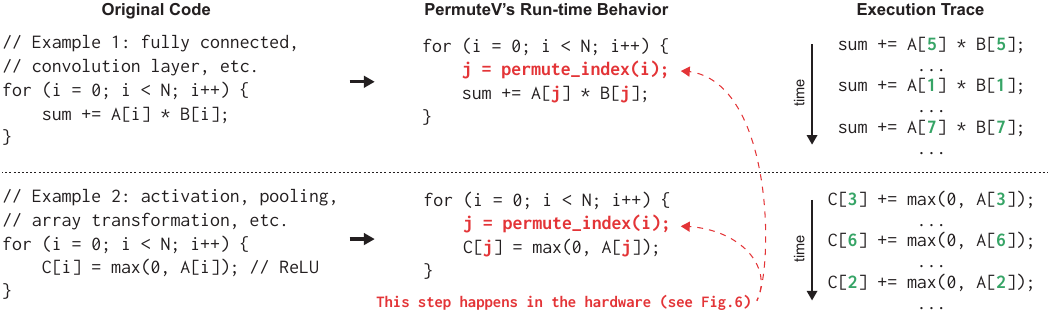}
    \caption{Our proposed defense philosophy: permute the execution order of loop iterations at run-time through our custom hardware module and ISA extension.}
    \label{figure:highlevel_pseudocode}
\end{figure*}

\subsection{RISC-V and the Ibex Core}

RISC-V \cite{7478332} is an open-source instruction set architecture (ISA) that is rapidly gaining attention in the semiconductor industry. Unlike proprietary ISAs like x86 and ARM, RISC-V is freely available, empowering designers to customize chips for specific applications. This openness fosters innovation and diversity, leading to a wide range of RISC-V cores from low-power micro-controllers to high-performance processors.

Ibex \cite{ibex} is an in-order, single-issue 32-bit RISC-V core with two pipeline stages designed for embedded control applications. It supports the Integer (I) or Embedded (E), Integer Multiplication and Division (M), Compressed (C), and B (Bit Manipulation) extensions. The Ibex core is widely adopted as a key component in the PULP \cite{PULP} and OpenTitan \cite{opentitan} platforms and is one of the most popular open-source RISC-V cores on GitHub (i.e., with over 1,100 stars).

\subsection{Threat Model}

These devices are often directly accessible to users and potential attackers, which makes them more vulnerable to diverse physical side-channel attacks. We consider electromagnetic (EM) side-channel attacks as a primary threat due to the following factors:

\begin{enumerate}
    \item \textit{Single-Tenant Environment:} These devices typically run bare-metal software or simple real-time operating systems (RTOS) like Zephyr \cite{Zephyr} or FreeRTOS \cite{FreeRTOS}. This limited software environment makes it infeasible to exploit side-channel attacks that rely on malicious software, such as cache-based or branch-predictor-based attacks \cite{Gerlach2023,ahmadi2021side,gonzalez2019replicating}.

    \item \textit{Physical Accessibility:} Edge devices are often deployed in physical locations where an attacker can easily gain access to the device and perform invasive measurements.
    
    \item \textit{Non-Invasive Nature of EM Attacks:} EM side-channel attacks are non-invasive, requiring only the proximity of a measurement device to the target device. This makes them a practical and feasible attack vector, unlike techniques like power analysis or voltage glitching that require physical modifications to the device.
\end{enumerate}

We adopt the same threat model as in the representative prior works~\cite{Batina2019,Takatoi2020,ShuffleV}, in which an attacker possesses the following capabilities: (i) the attacker has physical access to the target device; (ii) the attacker can manipulate the inputs to the device and observe its outputs; (iii) the attacker can utilize EM measurement equipment to capture the device's EM emanations.

\section{Our Proposed Solution: \ouralg}
We present the details of our proposed defense \ouralg by giving a high-level overview in Sec. \ref{subsection:overview}. Sec. \ref{subsection:loop-index-gen} provides detail of our custom hardware unit. Then, we explain our ISAs extension in Sec. \ref{subsec:isa-extension} and the compiler modification needed to utilize the hardware module in Sec. \ref{subsec:programming_support}.

\subsection{Overview} \label{subsection:overview}
We observe that the success of these existing EM-based SCAs \cite{Batina2019,Takatoi2020,ShuffleV} lies in their ability to perform statistical analysis of repetitive side-channel leakages associated with the runtime behavior of a victim application, e.g., the EM emanations caused by the MAC operation in an NN model. Inspired by this observation, we develop the \ouralg core to randomly shuffle the execution order of the loop iteration instead of executing the loop sequentially as in a normal processor. Consequently, the computation associated with a specific secret value (NN weight) will occur at a different time when the program is executed multiple times.
Therefore, the power and/or EM side channel attack is thwarted without any code modification. Fig. \ref{figure:highlevel_pseudocode} shows some example programs and equivalent codes illustrating the runtime behavior of our proposed defense philosophy.

Adhering to this defense methodology, we implement three key modifications in the core and compiler. (i) A \textit{Loop Index Generator (LIG) module} is added to the core to generate the permute index sequence. (ii) \textit{ISA Extension:} We extend the RISC-V ISA with a custom instruction to set up the LIG module and retrieve the permuted iteration index for the current iteration. (iii) \textit{Programming Support:} The compiler needs to be extended to emit our custom instruction to permute the loop. Several considerations are necessary to ensure correctness, e.g., the loop iteration must be independent. By co-designing both the hardware unit and software support, we can achieve low overhead and easy-to-use protection for software developers.

\subsection{Loop Index Generator} \label{subsection:loop-index-gen}

This section describes the design of our Loop Index Generator (LIG) module, which accepts the total number of loop iterations ($N$) as an input and outputs a sequence of permuted iteration indices (\texttt{permute\_index} in Fig.~\ref{figure:highlevel_pseudocode}). As a result, this new iteration index will be used instead in the loop body. For example, when $N$ = 8, the module may output 6, 3, 2, 0, ..., 7. Although intuitively feasible, there are two key challenges to overcome in realizing the loop index generator:

\begin{enumerate}
    \item \textit{Hardware size.} The module should support large $N$, e.g., 1,000 or higher, which are common in edge AI inference applications. However, the hardware for generating permutations also becomes larger when $N$ is large. %

    \item \textit{Memory locality.} Permuting the loop iteration may result in a random memory access pattern, which could impact the hardware performance in cores with complex memory hierarchies, such as cache and slow flash memory.
\end{enumerate}

\begin{algorithm}[b]
\caption{\ouralg pseudo-permutation strategy}
\label{alg:pseudo_permute}
\begin{algorithmic}[1]
\small
\State Let $N$ be the number of loop iteration, and $B$ be the block size (configured in the Loop Index Generator module)
\State $offset \gets randInt(0, N)$ \Comment{Random an int between $[0, N)$} 
\For{$i \gets 1$ to $\lceil N/B \rceil$}
    \State $index \gets [0, 1, ..., B-1]$
    \If{$i * B \leq N$} \Comment{Don't permute the last block if $N \nmid B$} 
        \State $index \gets permute(index)$
    \EndIf    
    
    \For{$j \gets 0$ to $B-1$} 
        \State $output \gets (offset + index[j])\mod N$
    \EndFor
    \State $offset \gets offset + B$
\EndFor
\end{algorithmic}
\end{algorithm}

\begin{figure}[]
    \centering
    \includegraphics[width=\linewidth]{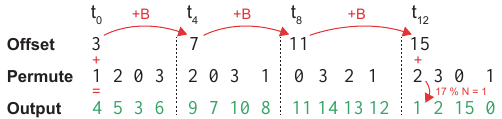}
    \caption{An example demonstrating \ouralg pseudo-permutation strategy when $N = 16$ and $B = 4$.}
    \label{figure:example_loop_index_generator}
\end{figure}

To address these problems, we proposed a \textit{pseudo-permutation strategy}, as described in Algorithm~\ref{alg:pseudo_permute}. The core idea is to permute the loop index within small, moving blocks of $B$ iterations, starting from a randomly selected iteration (offset). Specifically, we generate a permutation number within the range $[0, B-1]$, where $B$ represents the block size (configurable as a core parameter), then add an offset and modulo by $N$. This offset is initialized as a random value in the range $[0, N-1]$ and is incremented by $B$ every $B$ loop iterations. This combination produces the final permuted index. Fig. \ref{figure:example_loop_index_generator} illustrates how the algorithm works. First, we begin at iteration $3$ (i.e., assume $\text{offset} = 3$) and execute iterations $3$–$6$ in a permuted order, followed by iterations $7$–$10$, and so on. This strategy offers several advantages. First, it significantly reduces the hardware resources required to generate the permutation numbers. Second, it preserves spatial memory locality by ensuring access to nearby memory addresses in the consecutive $B$ cycles. Furthermore, it facilitates further optimization of the cache prefetching policy to accurately predict and fetch the subsequent block based on the offset value in the LIG module. More importantly, this approach achieves near-ideal randomness. Over many repetitive runs, the probability of each iteration, from $0$ to $N-1$, being executed in any given cycle ($t_0, t_1, \dots, t_N$) approaches $1/N$.

\begin{figure}[]
    \centering
    \includegraphics[width=0.85\linewidth]{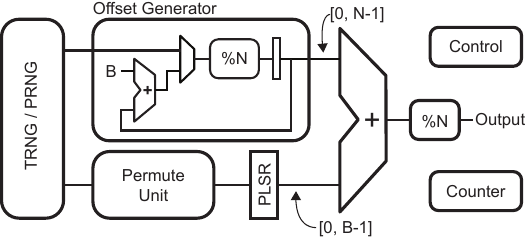}
    \caption{Block diagram of \ouralg's Loop Index Generator module.}
    \label{figure:block_diagram_loop_index_generator}
\end{figure}

We present the block diagram of the Loop Index Generator (LIG) hardware unit in Fig.~\ref{figure:block_diagram_loop_index_generator}. Upon initializing the module (using the \texttt{pv.init} instruction to set $N$ as described in Sec. \ref{subsec:isa-extension}), the offset generator produces an offset within the range $[0, N-1]$ by performing modulo by $N$ on a random number from the true-random number generator (TRNG) or pseudo-random number generator (PRNG). It updates the offset every $B$ loop iteration until the module is reinitialized with a new $N$ value. The permutation unit generates $B$ permutation numbers, which are then passed to a parallel-load shift register (PLSR) to produce one permuted number for each loop iteration. The final output is generated by adding the offset with the permuted number and modulo by $N$. To simplify the hardware design, the block size ($B$) is restricted to a power of two.

\begin{figure}[]
    \centering
    \includegraphics[width=0.9\linewidth]{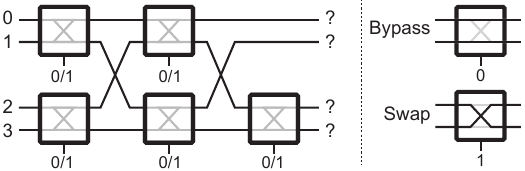}
    \caption{Overall design of the $2^n$ Permute Unit based on \cite{10.1145/321439.321449} when $B=4$.}
    \label{figure:block_diagram_permute_unit}
\end{figure}

Fig.~\ref{figure:block_diagram_permute_unit} illustrates the design of our permute unit based on the approach described in \cite{10.1145/321439.321449} using $B = 4$ as an example. The unit comprises several swap units, each controlled by a swap signal connected to individual bits of the random number generated by the random number generator. From our experiments, we found that a block size of $4$–$8$ is sufficient to effectively defend against EM SCAs as detailed in Sec. \ref{subsec:sidechannel_validation}.

\subsection{ISAs Extension} \label{subsec:isa-extension}

This section introduces \ouralg custom ISA extension to enable efficient access and computation using the permuted index. All instructions (except \texttt{pv.init(i)}) differ from its RV32 counterpart in that it implicitly combined the value read from source register 1 (\texttt{rs1}) with the current permuted index left shift by 0, 1, or 2. This design enables very efficient code generation as explained in Sec. \ref{subsec:programming_support} and allows us to provide \ouralg counterpart for 23 R/I/B-type instructions in RV32IM ISA (\texttt{add}, \texttt{sub}, \texttt{xor}, \texttt{or}, \texttt{and}, \texttt{sll(i)}, \texttt{srl(i)}, \texttt{sra(i)}, \texttt{slt(u)}, \texttt{mul(h/su/u)}, \texttt{div(u)}, \texttt{rem(u)}, \texttt{beq}, \texttt{bne}) with only a few multiplexers and adder added to the core datapath without introduce any new opcode and/or instruction format.

\begin{table}[]
\caption{List of \ouralg custom instructions and their descriptions. Ln.pi refers to the current permuted iteration number of LIG module \#n. Ln.i refers to the current iteration number of LIG module \#n. x refers to the shift amount (0, 1, or 2). }
\label{tab:instruction_description}
\scriptsize
\centering
\begin{tabular}{l|l}
\toprule
\textbf{Instruction}  & \textbf{Description} \\
\midrule
\multicolumn{2}{l}{\textbf{\ouralg Setup Instructions}} \\
\midrule
\texttt{pv.init~~ Ln, rs2} &  Set no. of iterations using register \\
\texttt{pv.initi~ Ln, imm} &  Set no. of iterations using immediate \\
\midrule
\multicolumn{2}{l}{\textbf{\ouralg Counterparts for RV32I R-type Instructions}} \\
\midrule
\texttt{pv.add~~~ Ln.x,rd,rs1,rs2} &  \texttt{rd=(rs1+(Ln.pi<<x))+rs2} \\
\texttt{pv.sub~~~ Ln.x,rd,rs1,rs2} &  \texttt{rd=(rs1+(Ln.pi<<x))-rs2} \\
\texttt{pv.xor~~~ Ln.x,rd,rs1,rs2} &  \texttt{rd=(rs1+(Ln.pi<<x))\^{}rs2} \\
\texttt{pv.or~~~~ Ln.x,rd,rs1,rs2} &  \texttt{rd=(rs1+(Ln.pi<<x))|rs2} \\
\texttt{pv.and~~~ Ln.x,rd,rs1,rs2} &  \texttt{rd=(rs1+(Ln.pi<<x))\&rs2} \\
\texttt{pv.sll~~~ Ln.x,rd,rs1,rs2} &  \texttt{rd=(rs1+(Ln.pi<<x))<<rs2} \\
\texttt{pv.srl~~~ Ln.x,rd,rs1,rs2} &  \texttt{rd=(rs1+(Ln.pi<<x))>>rs2} \\
\texttt{pv.sra~~~ Ln.x,rd,rs1,rs2} &  \texttt{rd=(rs1+(Ln.pi<<x))>>rs2} \\
\texttt{pv.slt(u)~Ln.x,rd,rs1,rs2} &  \texttt{rd=(rs1+(Ln.pi<<x))<rs2?1:0} \\
\midrule
\multicolumn{2}{l}{\textbf{\ouralg Counterparts for RV32I I-type Instructions}} \\
\midrule
\texttt{pv.slli~~ Ln.x,rd,rs1,imm} &  \texttt{rd=(rs1+(Ln.pi<<x))<<imm[4:0]} \\
\texttt{pv.srli~~ Ln.x,rd,rs1,imm} &  \texttt{rd=(rs1+(Ln.pi<<x))>>imm[4:0]} \\
\texttt{pv.srai~~ Ln.x,rd,rs1,imm} &  \texttt{rd=(rs1+(Ln.pi<<x))>>imm[4:0]} \\
\midrule
\multicolumn{2}{l}{\textbf{\ouralg Counterparts for RV32I B-type Instructions}} \\
\midrule
\texttt{pv.beq~~~ Ln.x,rs1,label} &  Branch if \texttt{rs1 $=$ (Ln.i<<x)} \\
\texttt{pv.bne~~~ Ln.x,rs1,label} &  Branch if \texttt{rs1 $\neq$ (Ln.i<<x)} \\
\midrule
\multicolumn{2}{l}{\textbf{\ouralg Counterparts for RV32M R-type Instructions}} \\
\midrule
\texttt{pv.mul~~~ Ln.x,rd,rs1,rs2} &  \texttt{rd=((rs1+(Ln.pi<<x))*rs2).lo} \\
\texttt{pv.mulh~~ Ln.x,rd,rs1,rs2} &  \texttt{rd=((rs1+(Ln.pi<<x))*rs2).hi} \\
\texttt{pv.mulsu~ Ln.x,rd,rs1,rs2} &  \texttt{rd=((rs1+(Ln.pi<<x))*rs2).hi} \\
\texttt{pv.mulu~~ Ln.x,rd,rs1,rs2} &  \texttt{rd=((rs1+(Ln.pi<<x))*rs2).hi} \\
\texttt{pv.div(u)~Ln.x,rd,rs1,rs2} &  \texttt{rd=(rs1+(Ln.pi<<x))/rs2} \\
\texttt{pv.rem(u)~Ln.x,rd,rs1,rs2} &  \texttt{rd=(rs1+(Ln.pi<<x))\%rs2} \\
\bottomrule
\end{tabular}
\end{table}

Tab.~\ref{tab:instruction_description} provides a list of \ouralg custom instructions for utilizing the LIG module and their description. The main difference between \ouralg instructions and their RV32IM counterparts is the addition of the \texttt{Ln.x} field, which is used to specify which LIG module to refer to (e.g., \texttt{L1}-\texttt{L3}) and the amount of left shift to apply to the read value (\texttt{x}). The \texttt{pv.init(i)} instruction accepts the total number of loop iterations and must be called before entering the loop to power on and initialize the LIG module. \ouralg's custom arithmetic and logic instructions, such as \texttt{pv.and}, \texttt{pv.slli}, and \texttt{pv.mul}, allow retrieval and computation using the permuted index from the LIG module in a single instruction. For instance, the \texttt{pv.add L1.2,t0,t1,t2} instruction adds the current permuted iteration number (\texttt{L1.pi}) left-shifted by 2 to \texttt{t1}, then adds the result with \texttt{t2} and stores the result in \texttt{t0}. By having the left-shifted version of the iteration number readily available, we can reduce the amount of instruction needed to calculate the load address or compare the loop terminal condition as explained in Sec. \ref{subsec:programming_support}. \ouralg branch instructions (\texttt{pv.beq} and \texttt{pv.bne}) compare \texttt{rs1} with the current non-permuted loop iteration number (\texttt{Ln.i}) left-shifted by \texttt{x} to simplify loop terminal condition checking. In addition, all \ouralg branch instructions automatically advance the internal counter and the permuted and non-permuted loop iteration number every time they are called.

\begin{figure}[]
    \centering
    \includegraphics[width=\linewidth]{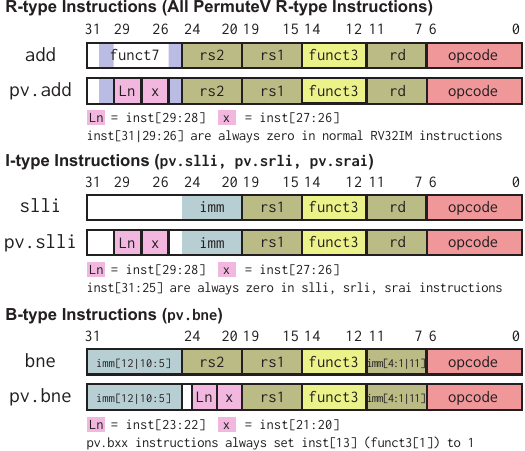}
    \caption{\ouralg's instruction encoding compared to their RV32I counterparts. The white color indicates bits that are unused in the original RV32IM instruction encoding.}
    \label{figure:instruction_format}
\end{figure}

Fig.\ref{figure:instruction_format} illustrates \ouralg custom instruction encoding compared to normal RV32 instruction encoding. The \texttt{Ln} field utilizes unused bits (always zero) in the corresponding RV32 instructions, allowing differentiation between normal RV32 instructions and \ouralg custom instruction easily by checking these bits. Two bits are reserved for the \texttt{Ln} field (bits 28-29 for R/I-type and bits 22-23 for B-type), allowing us to support up to 3 LIG modules in the core (\texttt{L1(01)} to \texttt{L3(11)}), i.e., up to 3 nested loops. Another two adjacent bits are also reserved to store \texttt{x} (left shift amount). This encoding maintains compatibility with standard RV32IM, allowing us to reuse the datapath and existing control signals, resulting in minimal hardware area overhead, as shown in Sec. \ref{subsec:hardware_area}.

\begin{figure*}[]
    \centering
    \includegraphics[width=0.9\linewidth]{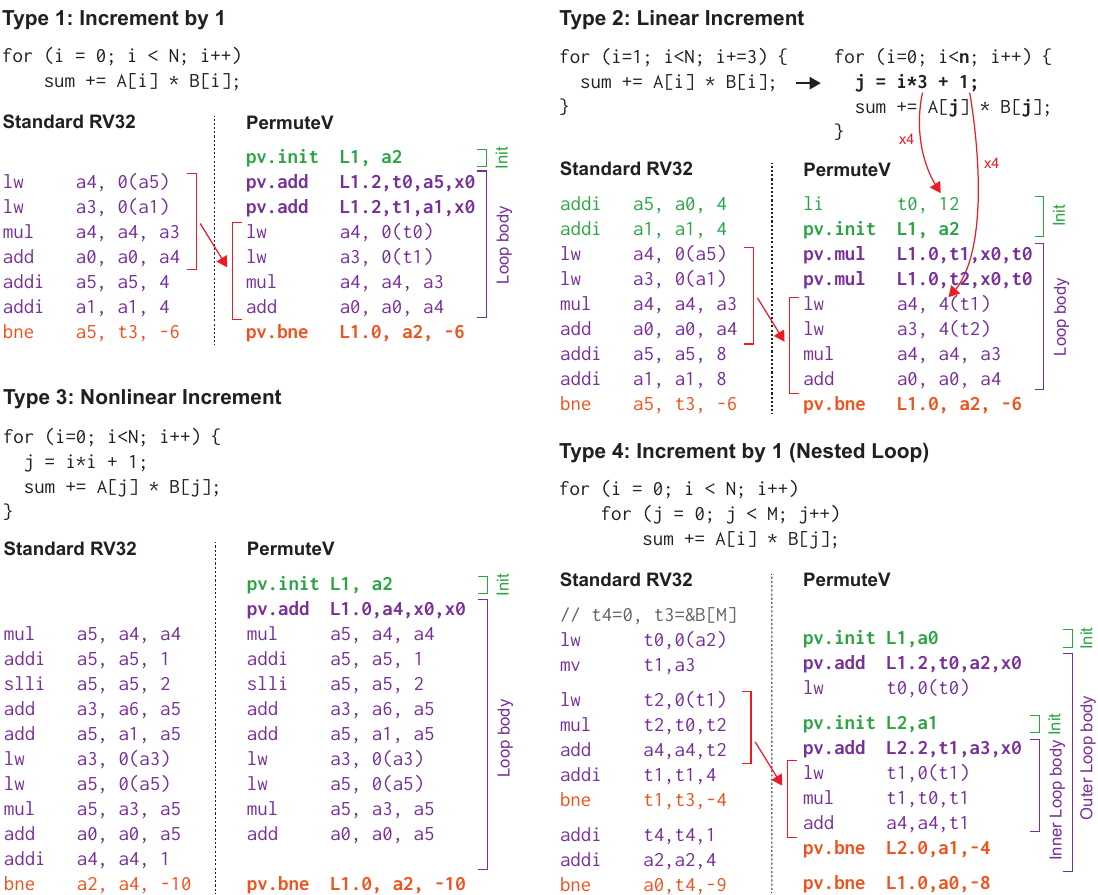}
    \caption{Example shows how to generate code for different loop styles using \ouralg's custom extension.}
    \label{figure:codegen_example}
\end{figure*}

\subsection{Programming Support} \label{subsec:programming_support}

Fig. \ref{figure:codegen_example} demonstrates how \ouralg's custom extension can be used to implement real-world loop code. \ouralg loop code begins by initializing the hardware module with the \texttt{pv.init/pv.initi} instruction using the module index (\texttt{L1-L3}) and the total number of iterations (\texttt{N}). We present sample codes for four common loop types.

\begin{enumerate}
    
\item  \textit{Increment by 1}. For a loop that increments by 1 (the most common case), the \texttt{pv.add} instruction can be used to calculate the load address by adding the current permuted iteration number (\texttt{L1.pi}) left-shifted by two (assuming each element consumes 1 word (4 bytes)) to the array-based address (\texttt{a5}), then adding the result with the zero register (\texttt{x0}) and storing the result in \texttt{t0} in a single instruction. The \texttt{pv.bne} is used instead of the normal \texttt{bne} instruction to compare the total number of iterations (\texttt{a2}) with the current non-permuted iteration number from the internal counter of the LIG module to continue or exit the loop.  

\item  \textit{Linear increment}. A loop with a linear increment includes all loops that perform increment by a fixed stride and can be thought of as a more generalized version of case 1. Thus, we can transform the loop to have a single canonical induction variable that starts at zero and steps by one. Then, we map the linear function calculating the actual index (e.g., \texttt{i*3+1}) to the \texttt{pv.mul} or \texttt{pv.slli} instruction (for power of two multiplier) and the offset field of the corresponding \texttt{lw} instruction. 

\item \textit{Nonlinear increment}. We retrieve the permuted iteration number using the \texttt{pv.add} instruction, passing zero as both source registers. Then, all address calculations are done the same way as in the standard RV32 code.

\item \textit{Nested loop}. A nested loop can be implemented using multiple LIG modules to permute both the outer and inner loops simultaneously. While we provide an example of increment by 1, nested loops with more complex conditions can be converted in the same way as non-nested loops.

\end{enumerate}

Note that in all cases the number of instructions in the loop body remains unchanged, enabling \ouralg to permute loop code with negligible runtime and code size overhead. 

\begin{figure*}[t]
    \centering
    \includegraphics[width=0.8\linewidth]{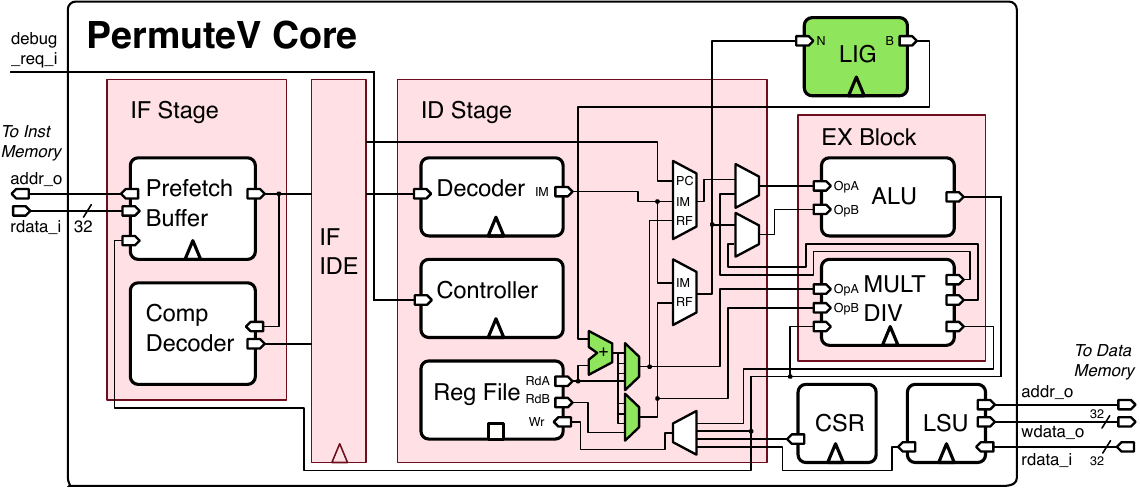}
    \caption{\ouralg core block diagram. New components added to the original Ibex core are highlighted in green.}
    \label{fig:blockdiagram}
\end{figure*}

To automate the following code transformation, we extends the LLVM compiler framework \cite{lattner2004llvm} with a new pass. The pass is executed on the Machine IR (MIR) following the instruction selection and prior to register allocation phase. Loop must pass the following criteria to be able to be transformed.

\begin{enumerate}
    \item \textit{Constant loop bound.} The loop bound must be constant for the duration of the loop e.g. $N$ in Fig. \ref{figure:codegen_example} must not be modified inside the loop body and there shouldn't be any \texttt{break} or \texttt{continue} statement inside the loop.

    \item \textit{Loop index.} The loop index must be mappable to linear or nonlinear continuous function at compile time e.g. there shouldn't be any branch which modify the loop index based on some runtime variables in the loop body. 

    \item \textit{Data dependency.} Each loop iteration must not depend on the result from other iterations with an exception of an associative and commutative reduction operation e.g. reduction sum in Fig. \ref{figure:codegen_example} which can be reordered without affecting the result.
\end{enumerate}

From our analysis, these conditions are met by the common loop code in NN inference. For example, fully connected and convolutional layers perform dot products, general matrix-vector multiplication (GEMV), or general matrix multiplication (GEMM), which are reduction-based operations. Activation layers (ReLU, tanh, etc.) scan and transform each element individually without any inter-iteration dependency.

\subsection{Hardware Implementation}

To facilitate the adoption of \ouralg in common design practice, we extend the open-source, widely used Ibex RISC-V core \cite{ibex}, an in-order, single-issue core with two pipeline stages. \ouralg is fully compatible with the Ibex core interface-wise, allowing it to serve as a drop-in replacement in many SoCs, e.g., those from OpenTitan \cite{opentitan} and PULP\cite{PULP}.

Fig.~\ref{fig:blockdiagram} illustrates the modifications made to the Ibex core. First, the LIG module was added to the ID/EX stage. The maximum number of LIG modules supported is parameterized and configurable from 1 to 3 (see Sec.~\ref{subsec:isa-extension}). To support all additional R/I-type instructions, we added an adder to compute \texttt{Reg[rs1] + Ln.i} and \texttt{Reg[rs1] + Ln.pi}, along with a multiplexer after the register file output to select between the adder results, the left-shifted version of the adder results, or the original \texttt{Reg[rs1]} value. For all new B-type instructions, one multiplexer is needed to select between the original \texttt{Reg[rs2]} value, the adder result, or the left-shifted version of the adder results. The select signal for both multiplexers is controlled by the \texttt{Ln} and the \texttt{x} field of the instruction. As all \ouralg custom instructions retain the same \texttt{opcode} and \texttt{funct3}/\texttt{funct7} values, no modifications to the instruction decoder or control signals are needed.

\subsection{RNG Selection}

We utilize the \texttt{systemc\_rng} pseudo-random number generator (PRNG) from OpenCores \cite{opencoresOverviewSystemCVerilog}, which integrates a Linear Feedback Shift Register (LFSR) with a Cellular Automata Shift Register (CASR) as described in \cite{Tkacik2002}. We synthesize this PRNG using a fixed initial seed for reproducibility and to establish a lower bound of \ouralg side-channel security in a less secure PRNG setup. Our assumption is that real-world implementation using a true-random number generator (TRNG) or a better PRNG should achieve higher security. Note that the design and evaluation of secure random number generators appropriate for FPGA and ASIC implementation is beyond the scope of this study. Nevertheless, the \ouralg is designed to be compatible with any RNG and is evaluated to be secure even with a simple PRNG, see Sec. \ref{subsec:sidechannel_validation}.

\section{Experimental Validation and Results}

\subsection{Experiment Setup} \label{subsec:general_experimental_setup}

\begin{figure}[tb]
    \centering
    \includegraphics[width=0.8\linewidth]{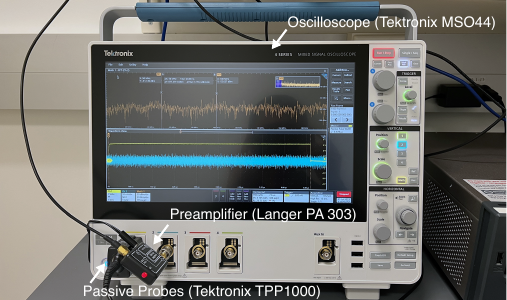}
    \includegraphics[width=0.8\linewidth]{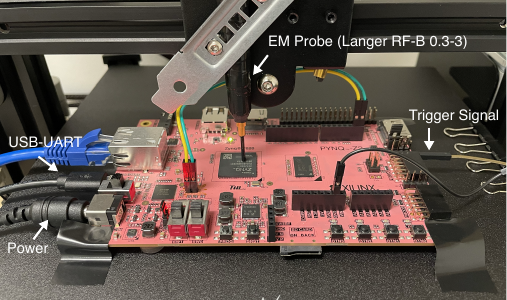}
    \caption{Experimental setup for performing EM measurements.} %
    \label{fig:ex_device}
\end{figure}

Our experimental setup, illustrated in Fig. \ref{fig:ex_device}, utilizes a Xilinx XUP PYNQ-Z2 FPGA board \cite{pynq-z2} equipped with a ZYNQ XC7Z020-1CLG400C System-on-Chip (SoC). All designs are synthesized using Xilinx Vivado Design Suite 2023.1 and operate at a clock frequency of 50 MHz. To capture EM side-channel leakage from the FPGA chip, we employ a Tektronix MSO44 4-BW-1000 mixed signal oscilloscope \cite{Oscilloscope} in conjunction with Langer H-field probes RF-B 0.3-3~\cite{prob} and pre-amplifiers PA-303~\cite{Preamplifier}.

\begin{table*}[]
\centering
\footnotesize
\caption{Number of traces required for a successful CEMA attack on multiply-accumulate (MAC) computation with vector length = 16 on Ibex \cite{ibex}, \ouralg ($B=4$), and \ouralg ($B=8$). \xmark indicates an unsuccessful attack at 1M traces. $^\dagger$ indicates attack successful only in a \textbf{white-box scenario with a very strong attacker}.}
\label{tab:cema_summary_num_trace}

\begin{tabularx}{\linewidth}{l|*{16}{Y}}
\toprule
\textbf{\backslashbox[20mm]{Core \xspace\quad}{Weight\#}} & \textbf{0} & \textbf{1} & \textbf{2} & \textbf{3} & \textbf{4} & \textbf{5} & \textbf{6} & \textbf{7} & \textbf{8} & \textbf{9} & \textbf{10} & \textbf{11} & \textbf{12} & \textbf{13} & \textbf{14} & \textbf{15} \\
\midrule
Ibex  & 213 & 22 & 65 & 91 & 498 & 198 & 134 & 224 & 236 & 644 & 513 & 282 & 103 & 1.1k & 121 & 167 \\
Our ($B=4$)  & 50k & \xmark & \xmark & \tbhl{101k} & \tbhl{587k} & \xmark & \tbhl{870k} & \tbhl{121k} & \xmark & \xmark & \tbhl{931k} & \tbhl{203k} & \xmark & \tbhl{490k} & \tbhl{59k} & \tbhl{1129} \\
Our ($B=8$)  & 5238 & \xmark & \xmark & \xmark & \xmark & \xmark & \xmark & \tbhl{54k} & \xmark & \xmark & \xmark & \xmark & \xmark & \xmark & \tbhl{37k} & \tbhl{2159} \\
\bottomrule
\end{tabularx}

\end{table*}

\subsection{Side-channel Validation} \label{subsec:sidechannel_validation}

We validate the effectiveness of \ouralg by conducting a CEMA attack using the Hamming weight leakage model to retrieve the vector value, during the dot product computation performing multiply-accumulate (MAC) operation on two vectors: A (user input) and B (pre-defined secret, e.g., NN weight), as shown in Fig. \ref{figure:highlevel_pseudocode}. Following \cite{Batina2019, Takatoi2020}, we choose the MAC operation to represent the NN confidentiality as it is the core operation formulating the convolutional and fully connected layers.

To guarantee a fair comparison across all core setups, we generated one million random input samples and used this same set consistently throughout the process of collecting electromagnetic (EM) traces. The attack itself involved feeding each unique input to the system and recording a corresponding EM trace. We consider two attack scenarios:

\begin{enumerate}
    \item \textit{White-box (i.e., strong attackers)}. In this scenario, we perform an attack on all weights, assuming that the actual weights (not permuted) prior to the attack position are known so that we can compute the expected Hamming weight of the summation result at the attack position.

    \item \textit{Black-box (i.e., real-world attackers)}.  In this scenario, we assume the attacker does not know the actual weight and relies on prior attack results to attack subsequent weights.
\end{enumerate}

\begin{figure}[]
    \centering
    \includegraphics[width=\linewidth]{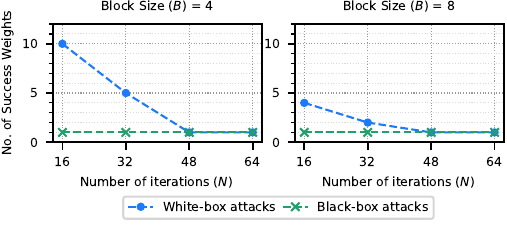}
    \caption{Number of weight that can be successfully attack vs. number of loop iteration on two \ouralg configuration ($B=4$ and $B=8$)}
    \label{figure:cema_num_weight_vs_num_iter}
\end{figure}

Fig. \ref{figure:cema_num_weight_vs_num_iter} presents the number of weights successfully recovered under EM SCA as the number of loop iterations ($N$) varies from 16 to 64, with block sizes ($B$) of 4 and 8. In the white-box attack scenario, we expected that the attack on the last weight would always work. This is due to the nature of reduction-based workloads (e.g., dot product), in which its final result (e.g., summation value in the last iteration) and its Hamming weight have to be the same regardless of the loop's permutation. This is not the case for a black-box attack, as the attack must be performed sequentially from the first weight onward and will likely fail if the attack on the prior weight is unsuccessful. Therefore, the minimum number of success weights should be at least 1 in the white-box scenario and 0 in the black-box scenario. Fig. \ref{figure:cema_num_weight_vs_num_iter} confirms this hypothesis, thereby validating the soundness of our attack methodology.

\textbf{Attack Complexity.} Tab. \ref{tab:cema_summary_num_trace} presents the CEMA results for a vector length of 16 (N=16). We show results when N=16, as it represents the most challenging scenario where \ouralg's protection is least effective. The results illustrate that the unprotected Ibex core can be successfully attacked with only a few hundred traces. In contrast, attacking \ouralg with a block size of 4 (\ouralg (B=4)) and a block size of 8 (\ouralg (B=8)) requires tens to hundreds of thousands of traces to extract 10 out of 16 and 4 out of 16 weights, respectively, in the white-box attack scenario. When considering the black-box scenario, only the first weight can be successfully attacked, demonstrating the effectiveness of our proposed solution in defense against EM SCA.

\textbf{Number of iteration ($N$) vs Side-channel Security.} 
Fig. \ref{figure:cema_num_weight_vs_num_iter} shows that the number of weights that can be successfully attacked is reduced significantly in both configurations when $N$ is higher. The trend is expected, as the probability of each weight being computed in its original index is roughly $1/N$. Therefore, the larger the value of $N$, the lesser the chance of a specific weight being computed at the same index across many runs, thus diminishing the CEMA attack success rate. 

\textbf{Block Size ($B$) vs Side-channel Security.} 
A larger block size makes our pseudo-random strategy behave closer to full permutation, but it will incur higher hardware area overhead. The impact of imperfect permutation is significant when $N$ is small. When the number of iterations is 48 or higher, both block sizes perform equally well. Therefore, using $B = 4$ is usually enough when considering NN inference workloads, which mostly have loop sizes in the range of at least a few hundred iterations.

\subsection{Hardware Area and Timing} \label{subsec:hardware_area}

We implement \ouralg as \ouralg Demo SoC by replacing the Ibex core in the Ibex demo system \cite{ibex_demo_soc} with the \ouralg core.
The Ibex demo system \cite{ibex_demo_soc} is a simple SoC design combining the Ibex core with some basic peripherals to support device programming, debugging, and interfacing with external devices.
Tab. \ref{tab:fpga_resource_timing} compares the resource utilization of \ouralg SoC with the Ibex Demo SoC and ShuffleV Demo SoC \cite{ShuffleV} which replaces the Ibex core in the Ibex Demo SoC with the Shuffle core, allowing for a fair comparison. When $B=4$, \ouralg core increases the FPGA resource usage with respect to the Ibex core by 270 look-up tables (+4.12\%) and 80 filp-flops (+1.36\%). The BRAM utilization is constant across all configurations, as it depends solely on the size of the program memory. In all configurations, \ouralg runs at the same speed as the baseline Ibex core \cite{ibex} at 50 MHz.

Since there is only one prior work on micro-architecture defense that targets NN inference, we provide a hardware area of the NN hardware inference engine (unmasked and masked version) \cite{Dubey2020,Dubey2020_ICCAD} as a reference. Although the results might not be directly comparable, they demonstrate that \ouralg can offer protection to NN inference at a significantly lower hardware area overhead (+6.42\% vs. +172.49\%).

\begin{table}[]
\caption{FPGA resource utilization of \ouralg vs. prior works. 'U' = unmasked (baseline), 'M' = masked (protect) implementation. '-' indicates value is not reported.}
\label{tab:fpga_resource_timing}
\centering
\footnotesize
\begin{tabularx}{\linewidth}{lYYY}
\toprule
\multicolumn{1}{c}{\textbf{Implementation}}    & \textbf{Device} & \multicolumn{1}{c}{\textbf{LUT}} & \multicolumn{1}{c}{\textbf{FF}} \\ 
\midrule
Ibex \cite{ibex}                    & PYNQ-Z2 & 6547                     & 5890    \\
\ouralg ($B=4$)                     & PYNQ-Z2 &  6817 (+4.12\%)          & 5970 (+1.36\%)    \\
\ouralg ($B=8$)                     & PYNQ-Z2 &  6967 (+6.42\%)          & 6024 (+2.27\%)     \\
ShuffleV (bs=4) \cite{ShuffleV}     & PYNQ-Z2 & 8079 (+23.4\%)           & 6915 (+17.4\%)   \\
\midrule
MaskedNet(U) \cite{Dubey2020} & Kintex-7 & 20296 & 18733  \\
MaskedNet(M) \cite{Dubey2020} & Kintex-7 & 55508 (+173.5\%) & 33290 (+77.7\%)  \\
\midrule
BoMaNet(U) \cite{Dubey2020_ICCAD} & Spartan-6 & 1833 & 1125  \\
BoMaNet(M) \cite{Dubey2020_ICCAD} & Spartan-6 & 9833 (+436.4\%) & 7624 (+577.7\%)  \\
\bottomrule
\end{tabularx}
\end{table}

\subsection{Performance} \label{subsec:performance}

\ouralg achieves negligible runtime overhead because the number of instructions within the loop body remains unchanged, as illustrated in Fig. \ref{figure:codegen_example}. While a few instructions may be added for loop initialization in some cases, this overhead is negligible compared to the time spent executing the loop itself. Furthermore, \ouralg pseudo-permutation strategy is designed to fully leverage prefetchers and caches if available. By operating on blocks of consecutive $B$ words (4–8 words) of memory (see Fig. \ref{figure:example_loop_index_generator}), the prefetcher can be utilized to automatically fetch the next $B$ words from the memory, in the same way as normal sequential codes. In addition, the actual memory accesses, while being permuted, are limited to the current block of $B$ words, which is likely to be kept in cache. This should lead to a high cache hit rate, similar to that of sequential code.

\section{Discussion and Future Work}

This section delves into the design trade-offs and limitations of \ouralg, proposing potential enhancements and outlining directions for future research.

\begin{enumerate}

    \item \textit{Memory locality.} Permuting the loop index can potentially impact memory locality and reduce performance. Nevertheless, our proposed pseudo-permutation strategy, detailed in Algorithm~\ref{alg:pseudo_permute}, mitigates this effect by permuting the loop index within a small sliding block of $B$ iterations. This approach preserves spatial memory locality by ensuring access to nearby memory addresses in consecutive $B$ cycles. 

    \item \textit{Compatibility with small loop.} As depicted in Fig. \ref{figure:cema_num_weight_vs_num_iter}, the efficacy of our proposed method diminishes for loops with a small number of iterations. However, this limitation is not a significant concern for real-world neural network inference, which typically involves loops with a large number of iterations. Nevertheless, this characteristic does restrict the applicability of \ouralg to certain cryptographic algorithms, such as AES, which can be implemented in software using loops with 16 iterations. Despite this, the prevalence of dedicated side-channel resistant hardware modules for common cryptographic algorithms like AES on most edge processors renders the support for software-based implementations of such algorithms less critical.

    \item \textit{Possibility for protecting other vulnerabilities.} Our threat model focuses on single-tenant embedded systems, which are not susceptible to software-based attacks like Spectre. However, we anticipate that \ouralg's non-deterministic execution and memory access pattern could extend its security benefits to mitigate timing and cache-based side-channel attacks. Further study will apply \ouralg to desktop-class processors to validate this claim.
\end{enumerate}

\section{Conclusion}
This paper introduces \ouralg, a RISC-V core designed to secure neural network inference on edge devices from electromagnetic (EM) side-channel attacks. By integrating a Loop Index Generator (LIG) module and custom ISA extensions, \ouralg implements a permutation strategy that effectively obfuscates EM emanations while maintaining efficient loop execution with negligible performance overhead. Our FPGA implementation demonstrates that \ouralg significantly enhances security against EM attacks, requiring orders of magnitude more traces for successful CEMA attacks, while maintaining minimal hardware overhead compared to the baseline Ibex core. Its compatibility as a drop-in replacement for the Ibex core ensures seamless integration into existing System-on-Chip (SoC) designs, broadening its applicability in edge AI scenarios. This work provides a practical and efficient solution to secure edge AI workloads, establishing a foundation for future research in scalable side-channel defenses.

\section*{Acknowledgements}
This work is supported in part by the U.S. National Science Foundation under Grants CNS-2153690, CNS-2247892, CNS-2239672, OAC-2319962, and CNS-2326597.

\bibliographystyle{IEEEtran}
\bibliography{references}

\end{document}